\documentstyle[psfig,12pt]{article}
\def\am{{$a_\mu$}}
\def\dam{{$\delta a_\mu$}}
\def\beq{\begin{equation}}
\def\eeq{\end{equation}}
\def\bea{\begin{eqnarray}}
\def\eea{\end{eqnarray}}
\def\ba{\begin{array}}
\def\ea{\end{array}}
\def\gappeq{\mathrel{\rlap {\raise.5ex\hbox{$>$}}
{\lower.5ex\hbox{$\sim$}}}}
\def\permil{$\%\raise.20ex\hbox{$_0$}}
\def\lappeq{\mathrel{\rlap{\raise.5ex\hbox{$<$}}
{\lower.5ex\hbox{$\sim$}}}}
\begin{document}
\topmargin -1.0cm
\oddsidemargin -0.8cm
\evensidemargin -0.8cm
\pagestyle{empty}
\begin{flushright}
CERN-TH/96-271 \\
DESY 96-211
\end{flushright}
\vspace*{5mm}
\begin{center}
{\Large \bf Constraints on Supersymmetric Models from the Muon
Anomalous Magnetic Moment}\\
\vspace{1.5cm}
{\large M. Carena$^{a,b}$,
G.F. Giudice$^a$\footnote{On leave of absence from INFN, Sezione di
Padova,
Padua, Italy.} and C.E.M. Wagner$^a$}\\
\vspace{0.3cm}
$^a$ Theory Division, CERN\\
Geneva, Switzerland\\
\vspace{0.3cm}
$^b$ Deutsches Elektronen-Synchrotronen, DESY\\
Hamburg, Germany\\
\vspace{0.3cm}
\vspace*{2cm}
Abstract
\end{center}
We study the impact of present and future $(g-2)_\mu$ measurements
on supersymmetric models.
The corrections to $(g-2)_\mu$ become particularly
relevant in the presence of light sleptons, charginos and neutralinos,
especially in the large $\tan\beta$ regime. For moderate or large
values of $\tan\beta$, it is possible to rule out scenarios
in which charginos and sneutrinos are
both light, but nevertheless escape
detection at the LEP2 collider. Furthermore,
models
in which supersymmetry breaking is transferred to the observable sector
through gauge interactions can be efficiently constrained by
the $(g-2)_{\mu}$ measurement.
\vfill
\begin{flushleft}
CERN-TH/96-271\\
DESY 96-211 \\
October 1996
\end{flushleft}
\eject
\pagestyle{empty}
\setcounter{page}{1}
\setcounter{footnote}{0}
\pagestyle{plain}

\section{Constraints from $a_\mu$}

The measurement of the anomalous magnetic moment of the muon \cite{pdg}
\beq
a_\mu\equiv \frac{g_\mu -2}{2}\equiv \frac{\mu_\mu}{(e\hbar /2m_\mu )}
-1 = (11~659~230 \pm 84 )\times 10^{-10}
\label{exp}
\eeq
has provided an extremely precise test to QED (for a review, see
\cite{mar} and references therein). The theoretical prediction for
\am\ in the context of the Standard Model has different contributions
which are usually divided into
\beq
a_\mu=a_\mu^{QED}+a_\mu^{EW}+a_\mu^{had}(vac~pol)+a_\mu^{had}(\gamma
\times \gamma )~.
\eeq

$a_\mu^{QED}$ contains the pure QED contribution which is known to order
$\alpha^5$ \cite{mar}. Extracting the
value of $\alpha$ from $g_e-2$ \cite{kin},
one obtains
\beq
a_\mu^{QED}=(11~658~470.6\pm0.2)\times 10^{-10}~.
\eeq

$a_\mu^{EW}$ contains the electroweak corrections which are now fully known
up to two loops \cite{loo}
\beq
a_\mu^{EW}=(15.1\pm0.4)\times 10^{-10}~.
\eeq

The largest source of uncertainty comes from the hadronic contributions,
which cannot be computed by perturbation theory alone.
$a_\mu^{had}(vac~pol)$ includes hadron vacuum polarization corrections which
enter at order $\alpha^2$. These corrections can be extracted from
$e^+e^-\to$ hadrons data by use of dispertion relations. A recent calculation
gives \cite{jeg}
\beq
a_\mu^{had}(vac~pol)=(702\pm 15)\times 10^{-10}~,
\eeq
in agreement with other modern evaluations (see ref.~\cite{jeg} for
a comparison of different estimates presented in the literature). Future
measurements of the cross sections for $e^+e^-\to$ hadrons at BEPC in
Beijing, at DA$\Phi$NE in Frascati, and at VEPP-2M in Novosibirsk are expected
to reduce the error in $a_\mu^{had}(vac~pol)$ to about $7\times 10^{-10}$
\cite{bar}.

The hadronic light by light amplitudes cannot be related to observables
and $a_\mu^{had}(\gamma
\times \gamma )$ must be estimated theoretically. Hayakawa {\it et al.}
\cite{hay}
give
\beq
a_\mu^{had}(\gamma
\times \gamma ) =(-5.2\pm 1.8)\times 10^{-10}~,
\label{haya}
\eeq
while Bijnens {\it et al.} \cite{bij} give
\beq
a_\mu^{had}(\gamma
\times \gamma ) =(-12.4\pm 5.0)\times 10^{-10}~.
\label{bijn}
\eeq
Summing the different contributions and combining errors in quadratures,
we obtain two theoretical estimates
for \am
\beq
a_\mu =(11~659~183 \pm 15)\times 10^{-10}
\label{theo1}
\eeq
\beq
a_\mu =(11~659~175 \pm 16)\times 10^{-10}~,
\label{theo2}
\eeq
depending on which of the two values of $a_\mu^{had}(\gamma
\times \gamma )$ in eqs.~(\ref{haya})-(\ref{bijn}) we use.

The experimental precision in the measurement of \am, presently at the
level of $84\times 10^{-10}$ is expected to be improved by
the E821 experiment at Brookhaven National Laboratory to the level of
$4\times 10^{-10}$, and possibly to 1--2$\times 10^{-10}$ if large
statistics is accumulated \cite{bun}. If theoretical errors in
$a_\mu^{had}$ are reduced, the E821 result will allow a direct test of the
electroweak corrections and therefore will also be sensitive to new
physics effects. Indeed already the present sensitivity can constrain
some new physics contributions, as it was shown in the case of
supersymmetry \cite{sup,nat}, light-gravitino interactions \cite{men},
compositeness \cite{com}, lepto-quarks \cite{lqs}, and light non-minimal
Higgs bosons \cite{pol}.

Constraints on new physics can be obtained by requiring that the new
contribution \dam\ lies within the difference between experimental
result and theoretical prediction. From eqs.~(\ref{exp}) and
(\ref{theo1})--(\ref{theo2}),
combining the theoretical and experimental errors in quadratures, we find
at 90\% C.L.
\beq
-90 \times 10^{-10} < \delta a_\mu < 190 \times 10^{-10}
\eeq
On the other hand, after the E821 experiment it will be possible to test
the value of $a_{\mu}$ at the level of $4\times 10^{-10}$.

In this paper we want to apply these constraints on \dam\ to specific cases
of interest, in the context of supersymmetric models.
We will first
show how, in a fairly model-independent way, \dam\ can rule out regions of
parameters with light charginos and sneutrinos which cannot be covered
by direct LEP2 searches. Then we will show how, in models with gauge-mediated
supersymmetry breaking, the bounds from \dam\ translate into strong
bounds on all supersymmetric particle masses.

\section{Chargino mass limits from LEP1.5 and LEP2}

The negative searches in the
LEP runs at 130 and 136 GeV have allowed to set a lower bound on the
chargino mass of 67.8 GeV \cite{ale},
if the chargino is gaugino-like and the sneutrino is heavy. This bound can be
relaxed in two cases we want to consider here. If the 
sneutrino is light and its mass is chosen appropriately,
the chargino production cross section suffers from a destructive interference
and it can be considerably smaller than in the case of heavy sneutrino.
The LEP1.5 limit on the chargino mass is then reduced,
especially if $\tan \beta$, the ratio of the
two Higgs vacuum expectation values, is large \cite{ale}.
The second case occurs when $m_{\chi^\pm}>m_{\tilde \nu}\gappeq
m_{\chi^\pm}-3$ GeV \cite{Aleph}. 
The chargino decay is then dominated by the two-body
decay $\chi^\pm \to \ell^\pm \tilde \nu$, but the final-state charged
lepton is too soft to be efficiently detected. In this case the LEP1.5
bound is completely lost, and the chargino could still be as light as
$m_Z/2$. This can happen in regions of parameters which cannot be excluded
by independent searches for neutralinos.
Notice that this problem will also remain in the LEP2 analyses at
higher $\sqrt{s}$. It is therefore important to understand if this
region of parameters, difficult for LEP searches, can be excluded by
other experiments. Recently the authors of ref.~\cite{ell} have argued
that this region of parameters can lead to an appropriate amount of
cold dark matter but cannot be excluded by cosmological constraints.
Here we want to study whether both regions where the LEP chargino limit
is reduced
can be excluded by the experimental data on \am.

As emphasized in ref.~\cite{nat} the supersymmetric
contributions to \am\ coming from smuon-neutralino and sneutrino-chargino
loops are significant and the present experimental bound already sets
important constraints on the parameters, especially if
$\tan \beta$ is large.
For $\tan \beta \gg 1$, the supersymmetric contribution is approximately
given by
\beq
\delta a_\mu \simeq \frac{\alpha}{8\pi \sin^2\theta_W}\frac{m_\mu^2}{\tilde
m^2}\tan\beta \simeq 15\times 10^{-10}\left(\frac{100~{\rm GeV}}{\tilde m}
\right)^2 \tan\beta ~,
\label{appr}
\eeq
where $\tilde m$ represents the typical mass scale of weakly-interacting
supersymmetric particles. It is evident from eq.~(\ref{appr}) that, if
$\tan \beta \gg 1$, the
experimental constraint on \dam\ can set bounds on the supersymmetric
particle masses which are competitive with the direct collider limits.
Indeed, the case $\tan \beta \simeq m_t/m_b \gg 1$ has some
special theoretical appeal.
First of all, it allows the unification of the bottom and tau Yukawa
couplings at the same energy scale at which gauge couplings 
unify, consistently with the prediction of the minimal $SU(5)$ GUT model.
Also it allows a dynamical explanation for the top-to-bottom mass ratio,
with approximately equal top and bottom Yukawa couplings at the GUT scale,
consistently with the minimal $SO(10)$ GUT \cite{yukawa}.

The supersymmetric contribution to \am\ is
\begin{eqnarray}
\delta a_{\mu}^{\chi^0} & = & \frac{m_{\mu}}{16 \pi^2} \sum_{mi}
\left\{
- \frac{m_{\mu}}{6 m_{\tilde{\mu}_m}^2 \left(1 - x_{mi}\right)^{4}}
\left(N_{mi}^L N_{mi}^L + N_{mi}^R N_{mi}^R \right)
\right.
\nonumber\\
& \times & \left( 1 - 6 x_{mi} + 3 x^2_{mi} + 2 x_{mi}^3 - 6 x_{mi}^2 \ln x_{mi}
\right)
\nonumber\\
& - &
\left.
\frac{m_{\chi^0_i}}{m_{\tilde{\mu}_m}^2 ( 1 - x_{mi})^{3}}
N_{mi}^L N_{mi}^R ( 1 - x_{mi}^2 + 2 x_{mi} \ln x_{mi}) \right\}
\end{eqnarray}
\begin{eqnarray}
\delta a_{\mu}^{\chi^+} & = & \frac{m_{\mu}}{16 \pi^2} \sum_k \left\{
\frac{m_{\mu}}{3 m_{\tilde{\nu}}^2 \left(1-x_k\right)^4}
\left(C_k^L C_k^L + C_k^R C_k^R \right)
\right.
\nonumber\\
& \times & \left( 1 + 1.5 x_k + 0.5 x_k^3
- 3 x_k^2 + 3 x_k \ln x_k \right)
\nonumber\\
& - &
\left.
\frac{ 3 m_{\chi^\pm_k}}{m_{\tilde{\nu}}^2 \left(1-x_k\right)^3} C_k^L
C_k^R \left( 1 - \frac{4 x_k}{3} + \frac{x_k^2}{3} + \frac{2}{3}
\ln x_k\right) \right\}
\end{eqnarray}
where $x_{mi} = m_{\chi^0_i}^2/m_{\tilde{\mu}_m}^2$, $x_k = m_{\chi^\pm_k}^2/
m_{\tilde{\nu}}^2$,
\begin{eqnarray}
N_{mi}^L & = & - \frac{m_{\mu}}{v_1} U^N_{3i} U^{\tilde{\mu}}_{Lm}
+ \sqrt{2} g_1 U^N_{1i} U^{\tilde{\mu}}_{Rm}
\nonumber\\
N_{mi}^R & = & - \frac{m_{\mu}}{v_1} U^N_{3i} U^{\tilde{\mu}}_{Rm}
- \frac{g_2}{\sqrt{2}}  U^N_{2i} U^{\tilde{\mu}}_{Lm} -
\frac{g_1}{\sqrt{2}}U^N_{1i} U^{\tilde{\mu}}_{Lm}
\nonumber\\
C_k^L & = & \frac{m_{\mu}}{v_1} U_{k2}
\nonumber\\
C_k^R & = & - g_2 V_{k1}
\end{eqnarray}
Here $U^N_{ij}$, $U^{\tilde{\mu}}_{(R,L)m}$, $U_{kl}$ and $V_{kl}$
are the neutralino, smuon and chargino mixing matrices,
$i,j=1,4$; $k,l=1,2$ and $m=1,2$;
$m_{\chi^0_i}$, $m_{\tilde{\mu}_m}$, $m_{\tilde{\nu}}$ and $m_{\chi^\pm_k}$
are the neutralino, smuon, sneutrino and chargino mass
eigenstates, $m_{\mu}$ is the muon mass and $g_i$ are
the electroweak gauge couplings.

The value of \dam\ depends on $M$, $\mu$
and $\tan \beta$ in the chargino
and neutralino sectors (we are assuming unification of gaugino masses),
and on the parameters $\tilde m_{L_L}$, $\tilde m_{E_R}$, $A$, which
determine the smuon mass matrix
\beq
m^2_{\tilde \mu}=\pmatrix{
{\tilde m}^2_{L_L}+m^2_\mu+
(-\frac{1}{2}+\sin^2
\theta_W)\cos 2\beta M_Z^2
          & m_\mu( A-\mu\tan\beta)\cr
m_\mu( A-\mu\tan\beta)&
{\tilde m}^2_{E_R}+m^2_\mu
-\sin^2\theta_W\cos 2\beta M_Z^2  }
\eeq
The parameter
$A$ appears only in the left-right smuon mixing, which is dominated by
the $\mu$ term in the large $\tan \beta$ region. As we will mainly focus to
this case, we can safely set $A=0$. Moreover the total result
is usually dominated by the sneutrino-chargino contribution, which is
independent of $A$. Finally the sneutrino mass square is given by
\beq
m_{\tilde \nu}^2={\tilde m}^2_{L_L}+\frac{1}{2}\cos 2 \beta M_Z^2
\eeq

As \dam\ is sensitive only to the mass of the muon sneutrino, while the
LEP1.5 bound is affected by any sneutrino, in order to proceed we have
to assume universality of sneutrino masses, $m_{\tilde \nu_e}=
m_{\tilde \nu_\mu}=m_{\tilde \nu_\tau}$. This hypothesis is
usually invoked to avoid unwanted lepton flavor violations (as in $\mu
\to e \gamma$) and it is satisfied by the minimal supersymmetric model.
It should be noticed however that this assumption is
not a necessary requirement for the suppression of flavor-changing neutral
current processes, as this can also be guaranteed by additional global
symmetries \cite{sei} or by a dynamical principle \cite{noi}, even in
presence of large mass splittings among squarks and sleptons with different
flavours.

Figure 1 shows the present experimental limit on the chargino mass,
as a function of the sneutrino mass \cite{Aleph} 
in the large $\tan\beta$ region.
A value of $\tan\beta = 20$ has been chosen in
the figure, but the mass bounds are stable under
changes of $\tan\beta$, for $\tan\beta \geq 10$.
For large values of the
sneutrino mass $m_{\tilde{\nu}} \gg M_Z$,
the chargino mass bound is close to the kinematical limit and
it is insensitive to
the sneutrino mass. However, for lighter sneutrinos,
the destructive
interference in the chargino production cross section causes
a reduction of the chargino mass bound.
The results for two different values of $\mu$ are displayed,
$\mu = -100$ GeV (shaded area) and $\mu = -500$ GeV (dark shaded
area). In the case
$\mu =-100$ GeV the chargino mass limit is reduced by more than
5 GeV, as
the sneutrino destructive interference effects are maximal for
$\mu \simeq - {\cal{O}}(M_2)$.

For $m_{\tilde \nu}< m_{\chi^\pm}$, the
two-body decay channel $\chi^\pm \rightarrow l^+ \tilde{\nu}$ is
kinematically accessible and its branching ratio becomes of
order one. However for small values of $m_{\chi^\pm} - m_{\tilde{\nu}}$, the
charged leptons are too soft to be detectable. This is the origin
of the gap in the chargino bound shown in fig. 1. One might expect 
this gap  to be covered by neutralino searches, especially in the
higgsino region, where the neutralino production cross section is
sizable. Instead, the light sneutrino insures that the next-to-lightest
neutralino predominantly decays into invisible final states,
$\chi^0_2\to \chi^0_1 \nu \bar \nu$. As an extreme case, three sneutrinos,
two neutralinos, and one chargino could be just above the LEP1 threshold,
but escape searches at LEP2.

For a given set of chargino parameters
and sneutrino mass,
we have chosen the value of $\tilde m_{E_R}$ which minimizes the effect
on \dam, in order to obtain the most conservative bound. This bound is
shown in fig.~1 and superimposed to the experimental limit.
The present constraint on \dam\ already closes the ``hole" left by LEP1.5,
if $\tan \beta$ is large enough. Indeed, for $|\mu| = {\cal{O}}$(100 GeV),
the hole, which would survive after the final LEP2 run if no chargino
is found, can be 
closed through the $\delta a_{\mu}$ constraints
for $\tan\beta \geq 10$ (20) for negative (positive)
values of $\mu$. For $|\mu| = {\cal{O}}$(500 GeV), for which the lightest
chargino and neutralino are mostly gauginos, $|\delta a_{\mu}|$ is
slightly suppressed, leading to somewhat weaker bounds. In this case,
the hole is closed for $\tan\beta \geq 20$ (40) for negative (positive)
values of $\mu$. Indeed, these bounds may be inferred from fig. 1,
by taking into account the approximate
linear dependence of $\delta a_{\mu}$ with
$\tan\beta$ in the large $\tan\beta$ regime and the fact that the sign of
$\delta a_{\mu}$ is given by the sign of the $\mu$ parameter.
The bounds become also somewhat weaker deep into the
Higgsino region ($M_{1/2} \gg M_Z$). The lightest neutralino becomes
almost degenerate in mass with the lightest chargino in this region,
and, indepedently on the sneutrino mass, no experimental limit may
be set if their mass difference is below 5 GeV \cite{chneu}.

In the above we have minimized $\delta a_{\mu}$ by scanning over the
right handed smuon mass, up to values of order 1 TeV. If, instead,
the value of the right handed smuon is restricted to be 
of order of the left handed one, for instance below 200 GeV,
the above results will be modified, depending on the gaugino and
Higgsino components of the light chargino and neutralinos. 
The dependence on the maximum right-handed smuon mass is significant
when $\mu$ is large. Indeed, in this case,
the chargino diagram contributions 
are suppressed and bino-exchange diagram provides the dominant
contribution. The contribution of this diagram is minimized
for large values of the right handed smuon mass.
Hence, more stringent
bounds than the ones obtained in the case of very heavy
smuons appear in this case. Numerically, the minimal value of
$\tan\beta$ for which the hole can be closed for $\mu \simeq -500$ GeV
changes from 20 to 14.
In the case of small $\mu$, $|\mu| = {\cal{O}}$(100 GeV), both
the chargino and neutralino contributions are relevant and a partial
cancellation takes place between them. In this case, the minimal value
of $\delta a_{\mu}$ is obtained for low values of the right handed
smuon mass and hence no variations in the previous bounds are obtained
by restricting the value of the right-handed smuon mass.

We also want to point out that a similar analysis can help in closing
``holes" in the LEP2 search for charged sleptons. In particular the
selectron
production cross section can vary by more than one order of magnitude
because of interference among the different contributions \cite{lep}.
This makes
the search harder, as the production rate can become very small.
Future LEP2 analyses can benefit from the \dam\ bound, as this narrows
considerably the allowed variation of the relevant parameters.

\section{Gauge-mediated supersymmetry breaking}

Theories with gauge-mediated supersymmetry breaking \cite{gms} have
recently received renewed attention \cite{at1}--\cite{mur}, because of
their property of naturally suppressing flavour violations.
These theories have also the attractive feature of predicting the
supersymmetric mass spectrum in terms of few parameters. Assuming that the
messenger particle which communicate supersymmetry breaking belong to
complete GUT multiplets, the gaugino and squark or slepton masses are
respectively
\beq
m_{\lambda_j}=k_j\frac{\alpha_j}{4\pi}\Lambda_G ~\left[
1+{\cal O}(F^2/M^4) \right]
\label{gaugi}
\eeq
\beq
{\widetilde m}^2=2\sum_{j=1}^3 C_jk_j\left( \frac{\alpha_j}{4\pi}\right)^2
\Lambda_S^2~\left[ 1+{\cal O}(F^2/M^4) \right]~,
\label{squak}
\eeq
where $k_1=5/3$, $k_2=k_3=1$, and $C_3=4/3$ for colour triplets, $C_2=3/4$
for weak doublets (and equal to zero otherwise), $C_1=Y^2$
($Y=Q-T_3$).
In the simplest models with minimal messenger structure, the gaugino and scalar
scales $\Lambda_G$ and $\Lambda_S$ are related by
\beq
\Lambda_G=\sqrt{n}\Lambda_S=n\frac{F}{M}~.
\label{enne}
\eeq
Here $M$ is the messenger mass scale and $\sqrt{F}$ is
the original
supersymmetry-breaking scale; $n$ is the effective number of messenger
fields. Perturbativity of gauge coupling constants up to the GUT scale
requires that the integer number $n$ satisfies $n\le 4$. The masses in
eqs.~(\ref{gaugi})--(\ref{squak}) are defined at the messenger mass scale
$M$,
and we have rescaled them to the physical mass value using the one-loop
renormalization group equations. We have chosen the messenger mass scale
$M=100$ TeV, but our results depend only mildly on this choice, because of
the slow logarithmic dependence. It is just
the ratio $F/M$ which really sets
the supersymmetric particle masses, and it will be taken as a free parameter
in our analysis. For a generic messenger sector, the energy scales
$\Lambda_G$ and $\Lambda_S$ are independent and hypercharge D-term
contributions can significantly affect the mass of the right-handed
smuon \cite{sme}. For simplicity we will restrict our consideration to
the minimal case in which eq.~(\ref{enne}) holds.

Besides the parameters $F/M$ and $n$ which describe the gaugino and scalar
spectrum, we also need to introduce the parameters $\mu$ and $\tan \beta$
which define the higgsino mass and mixings as in the ordinary supersymmetric
model considered in sect.~2. In gauge-mediated supersymmetric theories,
$\mu$ originates from new interactions beyond the usual gauge forces
\cite{mmu}, and its relation with $F/M$ depends on unknown constants.

We present our results of \dam\ as a function of $\mu$ and the weak
gaugino mass $M_2$ (which determines $F/M$), for fixed values of $n$
and $\tan \beta$. As discussed previously, the stronger limits from
\dam\ come in the region $\tan \beta \gg 1$. In this region, an important
constraint comes from the requirement that the determinant of the stau
square mass matrix is positive. In gauge-mediated supersymmetric theories,
the trilinear term $A$ vanishes at the messenger mass scale. Therefore
the slepton left-right mixing is dominated by the $\mu$ term, which
can become
dangerously large if $\tan \beta \gg 1$.

Figure 2 and 3 show the bounds which may be obtained in these models
in the $M_2-\mu$ plane for two different values of $\tan\beta$ and
for $n = 1$, 3, respectively. Large values of $|\mu|$ are restricted
by the lower bound on the stau mass,
while low
values of $M_{2}$ lead to unacceptable values of $\delta a_{\mu}$.
The bounds on $M_2$ become particularly strong when $\tan \beta$ is close
to its extreme value
$\tan\beta \simeq m_t(m_t)/m_b(m_t) \simeq 60$.
Notice that the limits become more stringent as $n$ is increased. Indeed,
as apparent from eq.~(\ref{enne}), for a given value of the gaugino mass,
larger $n$ correspond to lighter
sleptons, thus to larger contributions to \dam.
Because of the mass relations in eqs.~(\ref{gaugi})--(\ref{squak}),
a bound on $M_2$ can be easily translated into bounds on the various
supersymmetric particle masses. For instance, 
the gluino mass is $M_{\tilde g}\simeq$ (2.9.-- 2.5) $M_2$, 
where the
variation in  the ratio $M_{\tilde g}/M_2$ 
comes from the scale dependence of the gaugino masses. Analogously,
the right-handed squark mass is
$m_{\tilde q}\simeq$ (4.1--3.5) $M_2/\sqrt{n}$.

Future limits on \dam\ will put very strong constraints on models
of gauge mediated supersymmetry breaking. Indeed, the forseen
experimental sensitivity is of the order of the effects which are
obtained for values of the gluino mass $M_{\tilde{g}}$ as high as
1.4 TeV (3 TeV)
for n= 1 and $\tan\beta =$ 10 (60) respectively. For n =3, the
experimental sensitivity is of the order of the effects obtained
for $M_{\tilde{g}} \simeq$  2 TeV (4 TeV) respectively. 
Figure 4 shows the values of the gluino and right-handed squark masses
which can be tested assuming the bound $|\delta a_\mu |<4\times 10^{-10}$.
This is just
the future sensitivity of the E821 experiment. Of course the actual
bounds will depend on how much the theoretical error can be reduced,
and on the central values of the experimental measurement and the
theoretical prediction. The limits on the sparticle masses corresponding
to a bound on $\delta a_{\mu}$ different from $4 \times 10^{-10}$
can be  obtained from fig. 4 by noticing that, in the large
$\tan\beta$ regime, the supersymmetric contribution to 
$\delta a_\mu$ is proportional to $\tan\beta$.
The limits are obtained by minimizing the effect on \dam\ as
a function of $\mu$, for fixed values of $M_2$ and $\tan \beta$.
As seen in the figure even for values of $\tan\beta$ as low as one,
values of the gluino masses of order 450, 600 GeV may be tested
for gauge mediated supersymmetry breaking models with $n=1,3$ respectively.
Hence, $\delta a_{\mu}$
will represent a crucial test of these models, even for moderate values
of $\tan\beta$.

Recently it
has been argued \cite{nir} that,
in theories with gauge-mediated supersymmetry breaking,
large $\tan \beta$ could
be a natural option. This is because the different Higgs mass parameter
may arise at different order in perturbation theory, allowing therefore
a natural hierarchy which leads to large values of $\tan \beta$.
We have found here
that constraints from \dam\ strongly bound
the large $\tan \beta$ region in these theories, and future
measurements of \am\ will give a definite test of
the proposal in ref.~\cite{nir}. On the other hand, it should
be mentioned that the
motivation for large $\tan \beta$ coming from $b$--$\tau$ unification
is weakened in these theories. As shown in ref.~\cite{mur}, the
messenger particles slow down the running of $\alpha_s$ as the energy
scale is increased. This has the effect of enhancing the ratio $m_b/m_\tau$
at low energies, and therefore $b$--$\tau$ unification can be  achieved only
at the price of a low $\alpha_s(M_Z)$. This effect becomes more important
as $n$ increases. Present LEP determinations of $\alpha_s(M_Z)$ are
already cornering $b$--$\tau$ unification in gauge-mediated scenarios
\cite{mur}.

In gauge-mediated theories the original scale of supersymmetry breaking can
be rather low, of the order of 100 TeV. If this is the case, the gravitino
is very light, of the order of the eV. The Goldstino component of the
gravitino has couplings much stronger than gravitational ones. The contribution
to \dam\ from gravitino-smuon loops is \cite{men}
\beq
\delta a_\mu =\frac{G_N}{36\pi}\frac{ m_\mu^2}{m_{3/2}^2}{\rm Tr}{
m_{\tilde{\mu}}^2}~,
\label{grav}
\eeq
where $G_N$ is the Newton constant, $m_{3/2}$ is the gravitino mass, and
the trace is taken over the $2\times 2$ smuon square mass matrix. Relating the
gravitino mass to the original scale of supersymmetry breaking,
$m_{3/2}=F\sqrt{(4\pi/3)G_N}$, and using the expression of the smuon mass
in eq.~(\ref{squak}), we can write eq.~(\ref{grav}) in terms of the
messenger mass scale $M$:
\beq
\delta a_\mu = \frac{n}{2}\left( \frac{\alpha m_\mu}
{2\pi^2\sin 2\theta_W M} \right)^2 \sim n\left( \frac{100~{\rm
TeV}}{M}\right)^2 \times 10^{-19}~.
\eeq
This contribution is too small to give a significant constraint to the model.

In conclusion, we have studied how the measurement of the muon anomalous
magnetic moment constrains supersymmetric models in two different
scenarios. We have first discussed the case in which a light chargino
evades detection at LEP2 because of the presence of a light sneutrino.
Present bounds on \am\ can rule out this scenario if $\tan\beta$ is
sufficiently large and therefore provide an important tool complementary
to direct collider searches. Then we have considered the case of
theories with gauge-mediated supersymmetry breaking. Because of the
mass relations among sleptons, charginos, and neutralinos, the bound on
\dam\ gives a very definite constraint on the whole supersymmetric mass
spectrum. Future measurements on \am\ can conclusively test these
models, particularly  in the moderate and large $\tan\beta$ regions.

We thank M.~Schmitt and S.~Dimopoulos for useful discussions. We would
also like to thank the Aspen Center for Physics, where part of this 
work has been done, for its hospitality.

\def\ijmp#1#2#3{{\it Int. Jour. Mod. Phys. }{\bf #1~}(19#2)~#3}
\def\pl#1#2#3{{\it Phys. Lett. }{\bf B#1~}(19#2)~#3}
\def\zp#1#2#3{{\it Z. Phys. }{\bf C#1~}(19#2)~#3}
\def\prl#1#2#3{{\it Phys. Rev. Lett. }{\bf #1~}(19#2)~#3}
\def\rmp#1#2#3{{\it Rev. Mod. Phys. }{\bf #1~}(19#2)~#3}
\def\prep#1#2#3{{\it Phys. Rep. }{\bf #1~}(19#2)~#3}
\def\pr#1#2#3{{\it Phys. Rev. }{\bf D#1~}(19#2)~#3}
\def\np#1#2#3{{\it Nucl. Phys. }{\bf B#1~}(19#2)~#3}
\def\mpl#1#2#3{{\it Mod. Phys. Lett. }{\bf #1~}(19#2)~#3}
\def\arnps#1#2#3{{\it Annu. Rev. Nucl. Part. Sci. }{\bf
#1~}(19#2)~#3}
\def\sjnp#1#2#3{{\it Sov. J. Nucl. Phys. }{\bf #1~}(19#2)~#3}
\def\jetp#1#2#3{{\it JETP Lett. }{\bf #1~}(19#2)~#3}
\def\app#1#2#3{{\it Acta Phys. Polon. }{\bf #1~}(19#2)~#3}
\def\rnc#1#2#3{{\it Riv. Nuovo Cim. }{\bf #1~}(19#2)~#3}
\def\ap#1#2#3{{\it Ann. Phys. }{\bf #1~}(19#2)~#3}
\def\ptp#1#2#3{{\it Prog. Theor. Phys. }{\bf #1~}(19#2)~#3}

\vfill\eject

\begin{figure}
\centerline{  
\psfig{figure=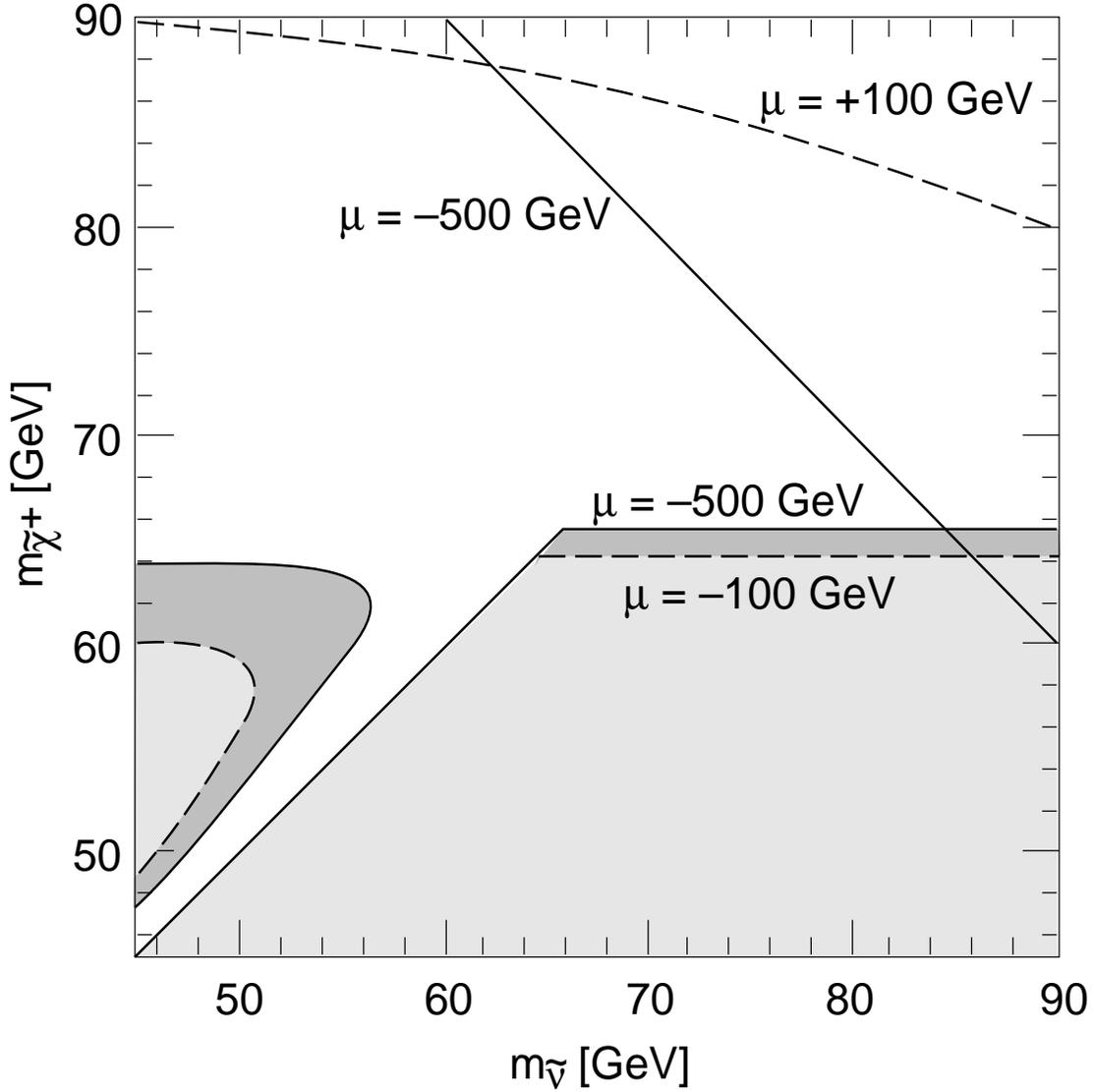,width=15cm,height=15.0cm}}
\caption{Present experimental limit on the 
lightest chargino mass as 
a function of the sneutrino mass $m_{\tilde{\nu}}$
for $\tan\beta = 20$, and two values of $\mu$,
$\mu =-500$ GeV (dark shaded region) and $\mu =-100$ GeV (light shaded
region). 
Also displayed in the figure are the present limits coming 
from constraints on
$\delta a_{\mu}$ for 
$\tan\beta=20$ and for $\mu = -500$ GeV and $\mu = 100$ GeV,
respectively.}
\label{fig1}
\end{figure}
\begin{figure}
\centerline{
\psfig{figure=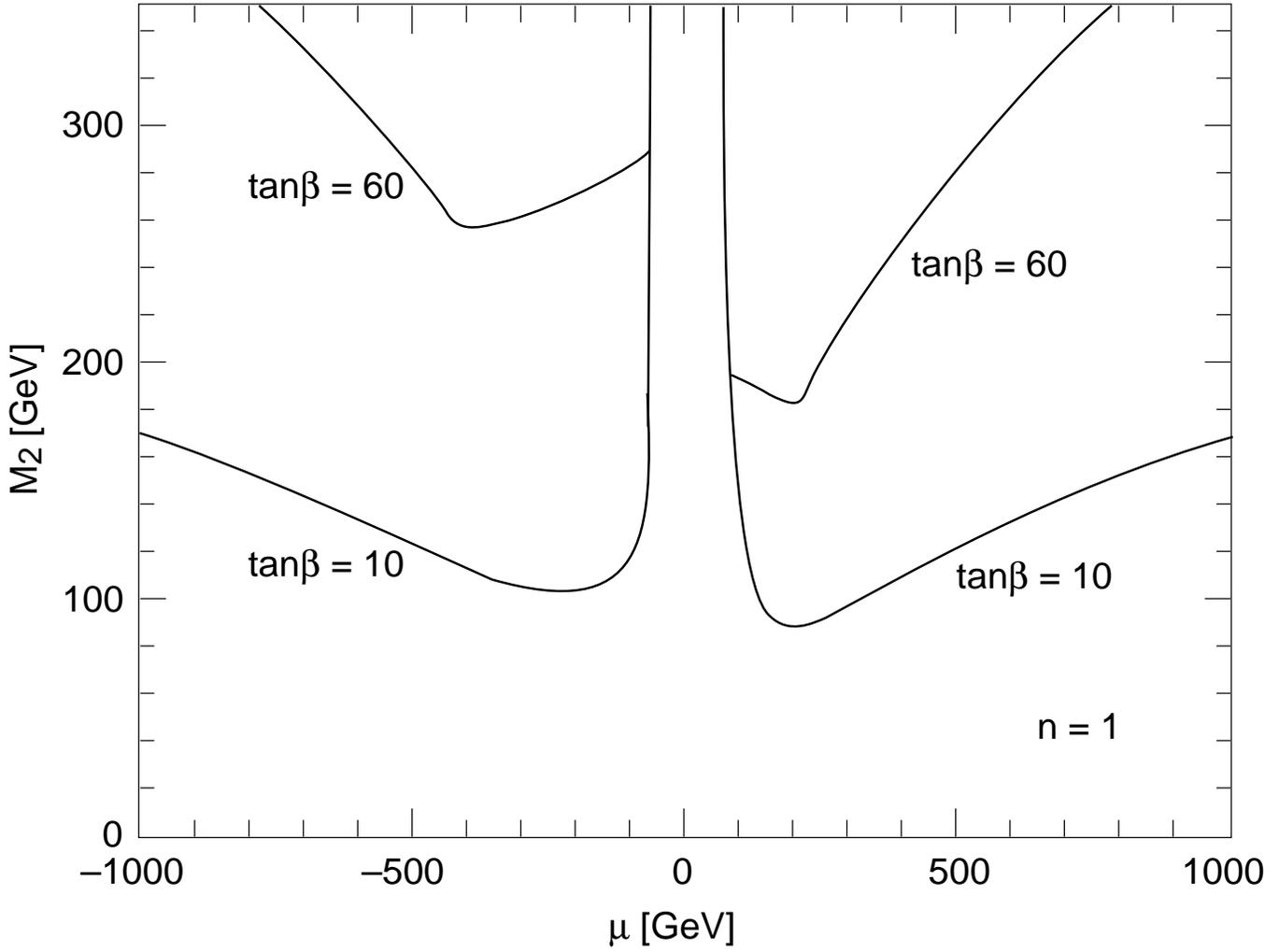,width=18cm,height=13.5cm}}
\caption{Present 
limits on the gaugino mass parameter 
$M_2$ as a function of the Higgsino mass parameter
$\mu$, in gauge mediated supersymmetry
breaking models with $n=1$. The upper curve represents the limit
for $\tan\beta = 60$, while the lower curve is the result for
$\tan\beta = 10$.}
\label{fig2}
\end{figure}
\begin{figure}
\centerline{
\psfig{figure=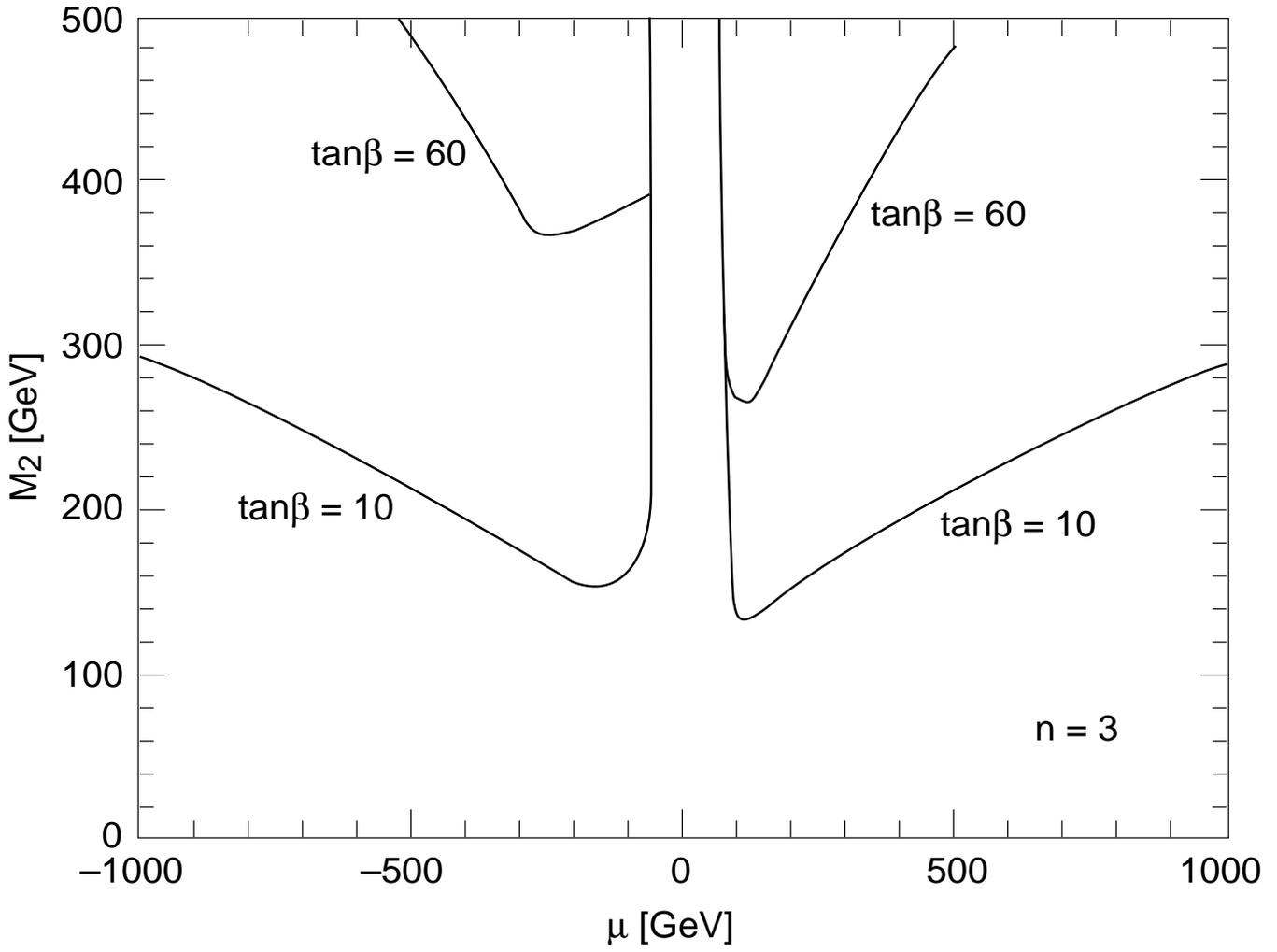,width=18cm,height=13.50cm}}
\caption{The same as fig. 2, but for $n=3$.}
\label{fig3}
\end{figure}
\begin{figure}
\centerline{
\psfig{figure=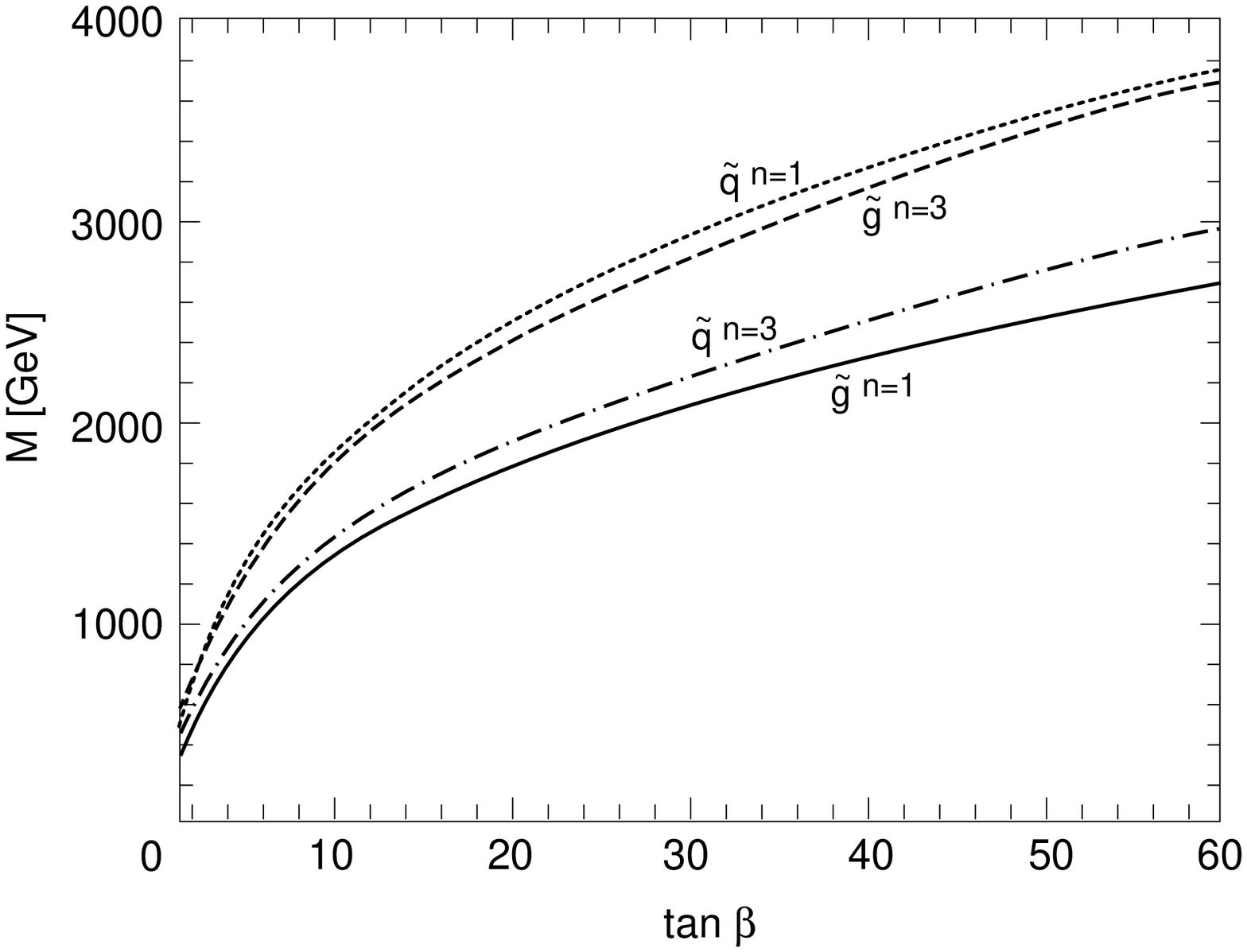,width=20cm,height=15.0cm}}
\caption{Limits on the gluino and right-handed squark masses as a function of
$\tan\beta$  for
gauge-mediated supersymmetry-breaking models with $n=1$ (solid and dotted
lines) and $n=3$ (dashed and dot-dashed lines), assuming
that the future  experimental sensitivity and the future theoretical
estimates will allow to constraints new physics effects at the
level $|\delta a_{\mu} |<
4 \times 10^{-4}$.}
\label{fig4}
\end{figure}

\begin{thebibliography}{99}

\bibitem{pdg} Particle Data Group, L. Montanet {\it et al.}
\pr{50}{94}{1171}.

\bibitem{mar} T. Kinoshita and W.J. Marciano, in {\it Quantum
Electrodynamics}, ed. T. Kinoshita (World Scientific, Singapore, 1990),
p. 419.

\bibitem{kin} T. Kinoshita, \prl{75}{95}{4728}.

\bibitem{loo} A. Czarnecki, B. Krause, and W.J. Marciano, \pr{52}{95}{2619}
and \prl{76}{96}{3267}.

\bibitem{jeg} F. Jegerlehner, in Proc. Workshop {\it ``QCD and QED in Higher
Orders"}, Rheinsberg, Germany, 1996, Nucl. Phys. B (Proc. Suppl.), to
appear.

\bibitem{bar} R. Barbieri and E. Remiddi, in {\it ``Second} DA$\Phi$NE
{\it Physics Handbook"}, eds. L. Maiani, L. Pancheri, and N. Paver (Frascati,
1995), Vol. II, p. 467;\\
P. Franzini, {\it ibid.} p. 471.

\bibitem{hay} M. Hayakawa, T. Kinoshita, and A.I. Sanda, \pr{54}{96}{3137}.

\bibitem{bij} J. Bijnens, E. Pallante, and J. Prades, preprint
NORDITA-95-75-N-P, hep-ph/9511388.

\bibitem{bun} G. Bunce, as quoted in ref. \cite{loo}

\bibitem{sup} P. Fayet, in {\it Unification of the Fundamental Particle
Interactions}, eds. S. Ferrara, J. Ellis and P. van Nieuwenhuizen
(Plenum Press, New York, 1980) p. 587;\\
J.A. Grifols and A. Mendez, \pr{26}{82}{1809};\\
J. Ellis, J.S. Hagelin and D.V. Nanopoulos, \pl{116}{82}{283};\\
R. Barbieri and L. Maiani, \pl{117}{82}{203};\\
D.A. Kosower, L.M. Krauss, and N. Sakai, \pl{133}{83}{305};\\
T.C. Yuan, R. Arnowitt, A.H. Chamseddine, and P. Nath, \zp{26}{84}{407}.

\bibitem{nat} J. Lopez, D.V. Nanopoulos, and X. Wang, \pr{49}{91}{366};\\
U. Chattopadhyay and P. Nath, \pr{53}{96}{1648};\\
T. Moroi, \pr{53}{96}{6565}.

\bibitem{men} A. Mendez and F.X. Orteu, \np{256}{85}{181}.

\bibitem{com} S. Brodsky and S. Drell, \pr{22}{80}{2236}.

\bibitem{lqs} G. Couture and H. K\"onig, \pr{53}{96}{555}.

\bibitem{pol} M. Krawczyk and J. Zochowski, preprint IFT-96/14,
hep-ph/9608321.

\bibitem{ale} D. Buskulic {\it et al.} (ALEPH Coll.), \pl{373}{96}{246};\\
M. Acciarri {\it et al.} (L3 Coll.), \pl{377}{96}{289};\\
P. Abreu {\it et al.} (DELPHI Coll.), preprint CERN-PPE-96-075;\\
G. Alexander {\it et al.} (OPAL Coll.), preprint CERN-PPE-96-096.

\bibitem{Aleph} D. Buskulic {\it et al.} (ALEPH Coll.),
preprint CERN-PPE/96-083.

\bibitem{ell} J. Ellis, T. Falk, K.A. Olive, and M. Schmitt, preprint
CERN-TH/96-102, hep-ph/9607292.

\bibitem{yukawa} 
M. Olechowski, and S. Pokorski, \pl{214}{88}{393};\\
B. Anantharayan, G. Lazarides, and Q. Shafi, \pr{44}{91}{1613};\\
G. Anderson, S. Dimopoulos, L.J. Hall, and  S. Raby, 
\pr{47}{93}{3702};\\
V. Barger, M.S. Berger, and P. Ohmann, \pr{47}{93}{1093};\\
M. Carena, S. Pokorski, and C.E.M. Wagner, \np{406}{93}{59};\\
L.J. Hall, R. Rattazzi, and U. Sarid, \pr{50}{94}{7048};\\
M. Carena, M. Olechowski, S. Pokorski, and C.E.M. Wagner, \np{426}{94}{269};\\
P. Langacker, and N. Polonsky, \pr{49}{94}{1454}.

\bibitem{sei} Y.~Nir and N.~Seiberg, \pl{309}{93}{337}.

\bibitem{noi} S.~Dimopoulos, G.F.~Giudice, and N.~Tetradis, \np{454}{95}{59}.

\bibitem{lep} {\it Physics at LEP2}, ed. by G. Altarelli {\it et al.},
CERN 96-01 (1996).

\bibitem{chneu} See, for example, S. Ambrosanio, M. Carena, B. Mele
and C.E.M. Wagner, \pl{373}{96}{107};\\
G.F. Giudice and A. Pomarol, \pl{372}{96}{253}.

\bibitem{gms}
M.~Dine, W.~Fischler, and M.~Srednicki, \np{189}{81}{575};\\
S.~Dimopoulos and S.~Raby, \np{192}{81}{353};\\
M.~Dine and W.~Fischler, \pl{110}{82}{227};\\
M.~Dine and M.~Srednicki, \np{202}{82}{238};\\
M.~Dine and W.~Fischler, \np{204}{82}{346};\\
L.~Alvarez-Gaum\'e, M.~Claudson, and M.~Wise, \np{207}{82}{96};\\
C.R.~Nappi and B.A.~Ovrut, \pl{113}{82}{175};\\
S.~Dimopoulos and S.~Raby, \np{219}{83}{479}.

\bibitem{at1} M.~Dine and A.E.~Nelson, \pr{48}{93}{1277};\\
M.~Dine, A.E.~Nelson, and Y.~Shirman,
\pr{51}{95}{1362};\\
M.~Dine, A.E.~Nelson, Y.~Nir, and Y.~Shirman, \pr{53}{96}{2658}.

\bibitem{at2} S. Dimopoulos, M. Dine, S. Raby and S. Thomas,
\prl{76}{96}{3494};\\
S. Ambrosanio, G.L. Kane, G.D. Kribs, and S.P. Martin, \prl{76}{96}{3498}.

\bibitem{mmu} G. Dvali, G.F. Giudice, and A. Pomarol, preprint CERN-TH/96-61,
hep-ph/9603238.

\bibitem{at3} S. Dimopoulos, S. Thomas, 
and J.D. Wells, preprint SLAC-PUB-7148,
hep-ph/9604452;\\
S. Ambrosanio, G.L. Kane, G.D. Kribs, and S.P. Martin,
preprint hep-ph/9605398;\\
.K.S. Babu, C. Kolda, and F. Wilczek, preprint IASSNS-HEP-96-55,
hep-ph/9605408;\\
S. Dimopoulos, G.F. Giudice, and A. Pomarol, preprint CERN-TH-96-171,
hep-ph/9607225;\\
S. Martin, preprint hep-ph/9608224;\\
S. Dimopoulos, S. Thomas, 
and J.D. Wells, preprint SLAC-PUB-7237,
hep-ph/9609434;\\
J.A. Bagger, K. Matchev, D.M. Pierce, and R.-J. Zhang, preprint
SLAC-PUB-7310, hep-ph/9609444.

\bibitem{sme} S. Dimopoulos and G.F. Giudice, preprint CERN-TH/96-255,
hep-ph/9609344.

\bibitem{nir} M. Dine, Y. Nir, and Y. Shirman, preprint SCIPP-96-30,
hep-ph/9607397.

\bibitem{mur} C.D.~Carone and H.~Murayama, \pr{53}{96}{1658}.

\end{thebibliography}
\end{document}